\documentclass[acmthm,nonacm=true,format=sigconf]{acmart}
\makeatletter
\def\@ACM@checkaffil{% Only warnings <<<<<<<<<<<<<<<<
    \if@ACM@instpresent\else
    \ClassWarningNoLine{\@classname}{No institution present for an affiliation}%
    \fi
    \if@ACM@citypresent\else
    \ClassWarningNoLine{\@classname}{No city present for an affiliation}%
    \fi
    \if@ACM@countrypresent\else
        \ClassWarningNoLine{\@classname}{No country present for an affiliation}%
    \fi
}
\makeatother

\usepackage{lipsum}

\usepackage[utf8]{inputenc}
\usepackage{subcaption}
\usepackage{booktabs}
\usepackage{ifdraft}
\usepackage{algorithm, algpseudocode}
\usepackage{xcolor}
\usepackage{xspace}
\usepackage{braket}
\usepackage{soul}
\usepackage[capitalize]{cleveref}
\usepackage{balance}
\usepackage{enumitem}
\usepackage{svg}
\usepackage{hyperref}

\newcommand{\ta}{Tumult Analytics\xspace}

\title{Tumult Analytics: a robust, easy-to-use, scalable, and expressive framework for differential privacy}

\author{
    Skye Berghel,
    Philip Bohannon,
    Damien Desfontaines,
    Charles Estes,
    Sam Haney,\\
    Luke Hartman,
    Michael Hay,
    Ashwin Machanavajjhala,
    Tom Magerlein,\\
    Gerome Miklau,
    Amritha Pai,
    William Sexton,
    Ruchit Shrestha
}
\date{}
\affiliation{
    \vspace{0.3cm}
    Tumult Labs\\
    science@tmlt.io
    \vspace{0.3cm}
}
\begin{document}

\setlist[itemize]{topsep=5pt}

\begin{abstract}
In this short paper, we outline the design of \ta, a Python framework for differential privacy used at institutions such as the U.S. Census Bureau, the Wikimedia Foundation, or the Internal Revenue Service.
\end{abstract}

\maketitle

\pagestyle{plain}

% This file contains all the sections inside the paper, so we don't have to keep the content in sync between .tex files using different templates.

\section{Introduction and design goals}
\label{sec:intro-design-goals}

Tumult Analytics is an open-source framework for releasing aggregate information from sensitive datasets with differential privacy%
\footnote{Readers unfamiliar with differential privacy~\cite{dwork2006calibrating} can refer to \cite{wood2018differential,desfontainesblog20210927} for non-technical introductions to this field, and \cite{dwork2014algorithmic,vadhan2017complexity} for more in-depth surveys.}
(DP).
It supports many standard operations (e.g. filters, joins, maps) and aggregations (e.g. counts, averages, quantiles).
It is currently used at institutions such as the IRS, the Wikimedia Foundation, and the U.S. Census Bureau.

Tumult Analytics is designed to satisfy the following desiderata.
\begin{itemize}
  \item \textbf{Robustness.} An analyst with access to the raw data, who wants to publish a differentially private version of it, can confidently use the platform and obtain the desired privacy guarantees.
  \item \textbf{Ease of use.} Any data scientist or engineer can successfully apply DP to their own data, possibly after using the platform and its documentation to learn about the necessary concepts. No expert-level math knowledge or in-depth understanding of DP theory is ever required.
  \item \textbf{Scalability and performance.} Tumult Analytics can run DP computations on arbitrarily-sized datasets. It uses computational resources that are on the same order of magnitude as the non-private version of the queries it evaluates.
  \item \textbf{Expressiveness.} Tumult Analytics is sufficiently feature-rich to power real-world use cases. Additionally, the underlying privacy accounting framework is extensible enough to support the addition of new data transformation operators, aggregate functions, or even privacy notions and accounting methods, without requiring deep design changes.
\end{itemize}

%Figure~\ref{fig:analytics-example} is an example of a Tumult Analytics program, which we discuss in more detail in Section~\ref{sec:analytics}.

\begin{figure*}[!ht]

\centering

%\includesvg[inkscapelatex=false,width=14cm]{analytics-example.svg}
\includegraphics[width=15.8cm]{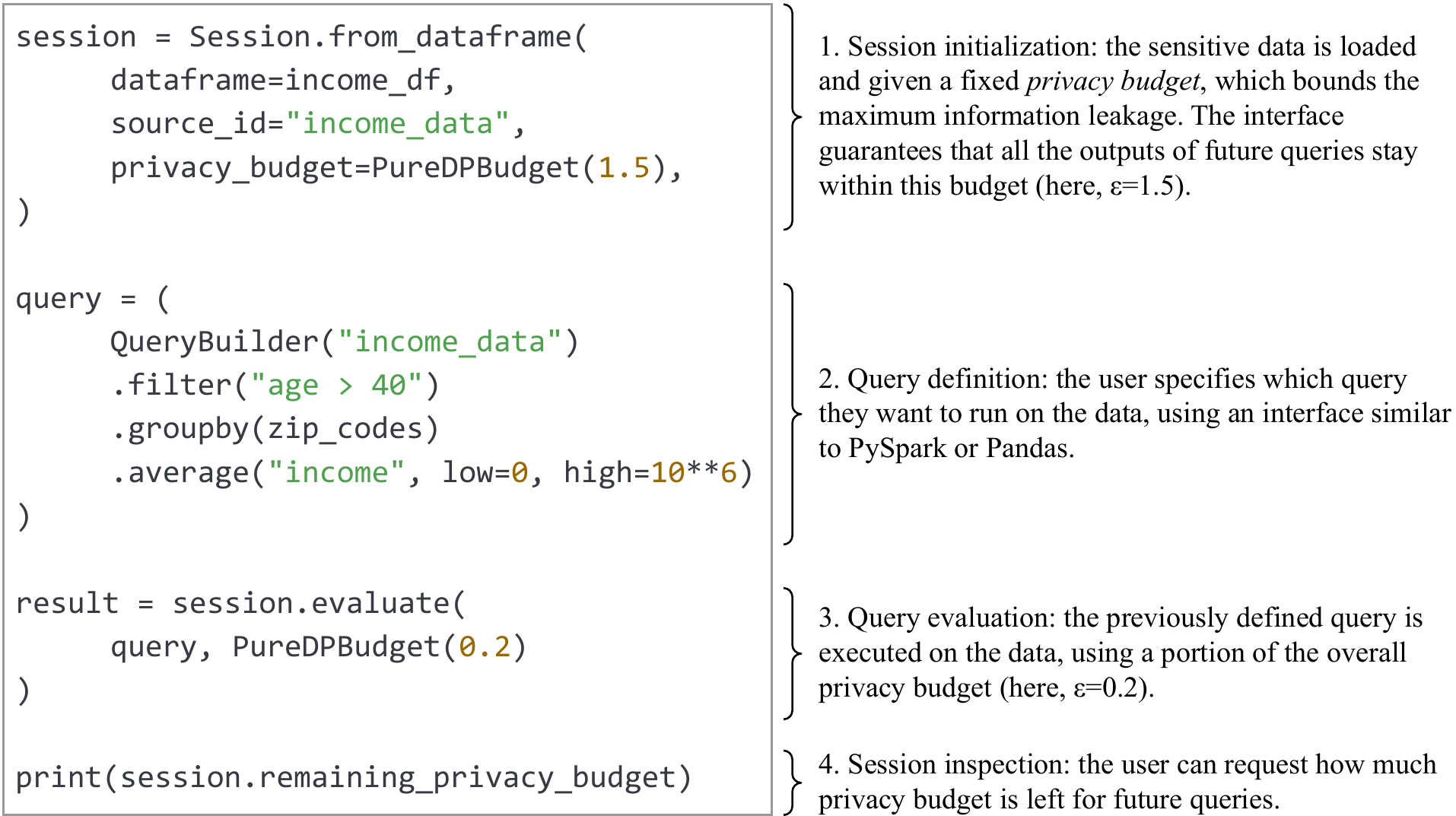}
% Source: https://docs.google.com/drawings/d/1PXO0BzxG_WLr1L2EmHEvmQoLDh4s6Q9W4d6vTfFHUqg/edit
\caption{An annotated Tumult Analytics program that computes the average income of all people older than 40 in an input dataframe \texttt{income\_df}, grouped by ZIP code.}
\label{fig:analytics-example}

\bigskip

%\includesvg[width=14cm]{core-diagram.svg}
\includegraphics[width=15.8cm]{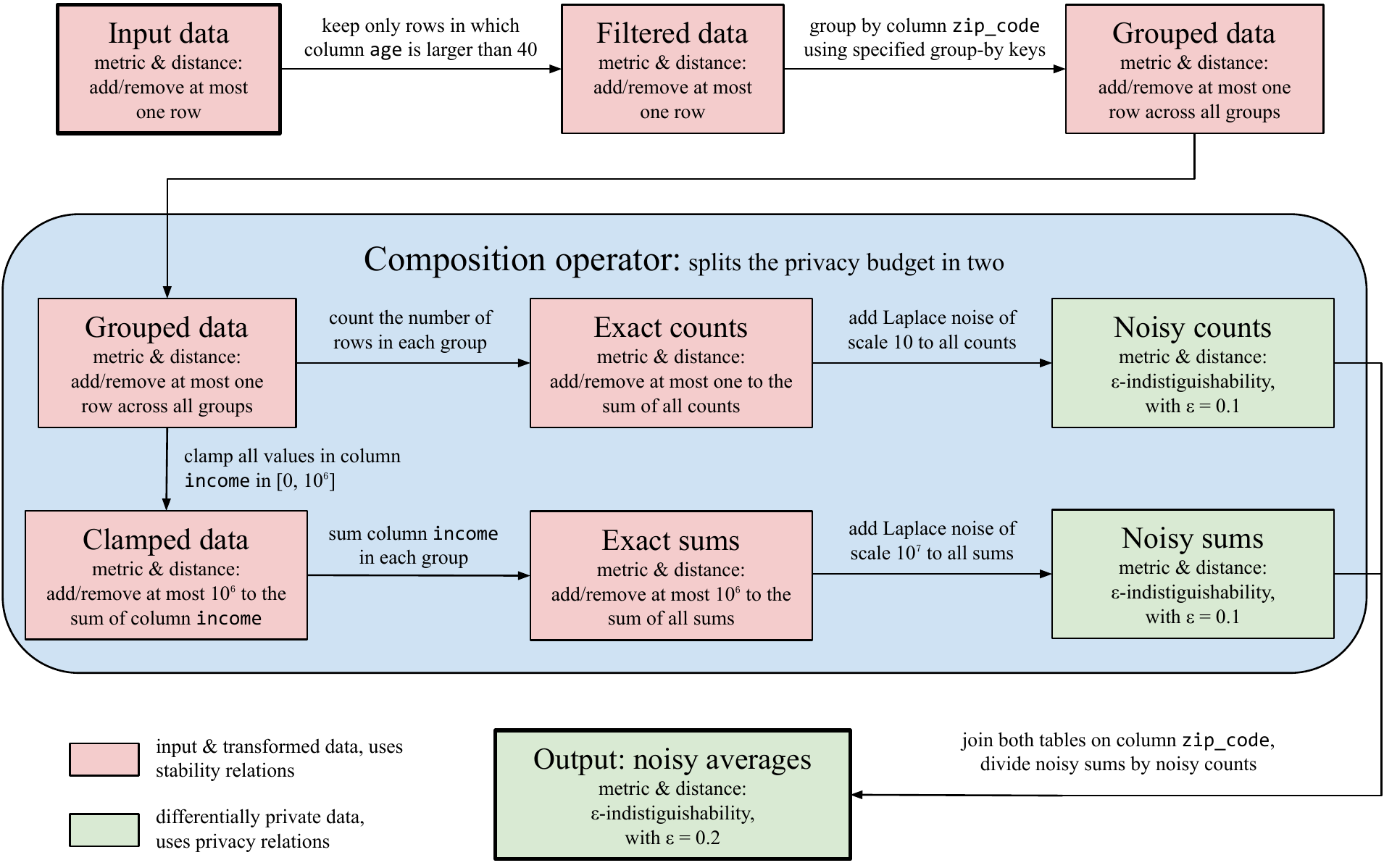}
% Source: https://docs.google.com/drawings/d/1LQpX_fQK1kijtqM3fxD02Z9IIy5iJhbFvk_DvvaCMsk/edit
\caption{A simplified diagram outlining the steps taken by the data through Tumult Core transformations, measurements and operators, to implement the query in Figure~\ref{fig:analytics-example}. Each labeled arrow corresponds to a Core component. The actual series of Core components to implement this query is more complex and incorporates some additional optimizations.}
\label{fig:core-diagram}

\end{figure*}

%\begin{lstlisting}[language=Python, caption=An example program for performing a private computation using Tumult Analytics. \label{lst:analytics-example}]
%session = Session.from_dataframe( (*@\label{line:session-creation-begin}@*)
%    dataframe=private_data,
%    source_id="my_data",
%    PureDPBudget(1.5)
%) (*@\label{line:session-creation-end}@*)
%query = ( (*@\label{line:query-begin}@*)
%    QueryBuilder("my_data")
%    .filter("age > 42")
%    .groupby(zip_codes)
%    .median("income", low=0, high=10**6)
%) (*@\label{line:query-end}@*)
%result = session.evaluate( (*@\label{line:evaluate-begin}@*)
%    query, PureDPBudget(0.2)
%)(*@\label{line:evaluate-end}@*)
%print(session.remaining_privacy_budget()) (*@\label{line:check-budget-begin}@*)
%# prints PureDPBudget(1.3) (*@\label{line:check-bugdet-end}@*)
%
%\end{lstlisting}

%Expressiveness and ease-of-use can sometimes be at odds, which is why the library is built in two separate components.
The library is built in two separate components.
\begin{itemize}
  \item Tumult Core is a privacy foundation, designed to be modular and extensible. We present its design in Section~\ref{sec:core}.
  \item Tumult Analytics focuses on ease-of-use, by providing a simpler interface on top of Tumult Core. We present its interface in Section~\ref{sec:analytics}.
\end{itemize}
We then compare Tumult Analytics with other open-source frameworks for differential privacy (Section~\ref{sec:comparison}).

\section{Tumult Core}
\label{sec:core}

Tumult Core is the underlying framework that Tumult Analytics relies on.
It is a collection of components and composition operators which allow users to implement complex differentially private mechanisms.
Tumult Core is where \emph{all} the privacy-critical logic lives: Tumult Analytics is only an interface to Tumult Core, and the underlying framework provides end-to-end privacy guarantees.
%We believe this framework is significantly more expressive than typical stability-based accounting schemes (see, e.g., \todo{pinq paper}).

The fundamental component of Tumult Core is called a \emph{measurement}.
A measurement takes an input from some domain and produces a randomized output.
The privacy properties of the measurement are captured by its \emph{input metric}, a function defining distances
%\footnote{Although technically, these distances need not satisfy the triangle inequality, and do not even need to be numbers -- here, ``distances'' are any set with a partial ordering.}
 on its input domain, its \emph{output measure}, a function defining distances between probability distributions over the output domain, and its \emph{privacy function}, a function that takes a distance in the input metric and returns a distance in the output measure.
A measurement $M$ with privacy function $f$ satisfies the following property: for any pair of inputs $x,y$ with distance at most $d$ in the input metric, the distributions $M(x)$ and $M(y)$ have distance at most $f(d)$ in the output measure.

This guarantee generalizes the usual DP guarantee in two ways.
First, it gives users flexibility to define alternative notions of neighbors.
In the usual definition of DP, neighbors are pairs of datasets that differ in a \emph{single} record:
it prevents an adversary from inferring information about a single record.
However, we often want a weaker guarantee -- for example, on the precise location of some person -- or a stronger guarantee -- for example, on all records associated with some individual.
Tumult Core measurements are general enough to capture these examples, as well as other similar variants (e.g. the N section in~\cite{desfontaines2020sok}).

Second, the privacy notion is abstracted in Tumult Core:
%Second, Tumult Core can represent different ways of measuring the privacy loss.
%In DP, if we run a mechanism $M$ on a pair of neighboring datasets $x$ and $y$, then $M(x)$ and $M(y)$ should be $\eps$-indistinguishable~\cite{dwork2006calibrating}. %, defined as
%
%\begin{equation}
%  D_{\infty}(M(x) \| M(y)) = \ln\left( \frac{\Pr[M(x) = S]}{\Pr[M(y) = S]} \right) \le \eps
%\end{equation}
%
%for all possible outputs $S$.
the framework is generic enough to cover not only differential privacy, but variants of DP as well, and potentially entirely new privacy notions.
Today, in addition to DP, Tumult Core supports zero-concentrated DP~\cite{bun2016concentrated}, and support for approximate DP~\cite{dwork2006our} and approximate zCDP is underway.
Other notions like R\'enyi DP~\cite{mironov2017renyi}, $f$-DP~\cite{dong2019gaussian}, and others (e.g. the Q section in~\cite{desfontaines2020sok}) could easily be supported.

In addition, Tumult Core also includes pre-processing components called \emph{transformations}.
Transformations include typical data processing operations (filters, maps, joins, \dots) and pre-processing building blocks (clamping, truncation, \dots) useful for DP algorithms.
Transformations don't have a privacy guarantee on their own.
Instead, they have a \emph{stability} guarantee, which relates a distance in the transformation's input metric to a distance in its output metric.

A Tumult Core program uses \emph{composition operators} to combine transformations and measurements, and implement complex algorithms.
When multiple components are combined, the stability or privacy function of the resulting component is inferred inductively.
This allows Tumult Core to build an end-to-end proof of the privacy guarantee of an entire program, no matter how complex.
Figure~\ref{fig:core-diagram} presents a simplified example of how the Tumult Analytics query in Figure~\ref{fig:analytics-example} is implemented in Tumult Core. 
The generality of this framework allows it to integrate various complex features, including:
\begin{itemize}
  \item \textbf{Private joins.} Private joins are notoriously difficult~\cite{kotsogiannis2019privatesql,dong2022r2t,johnson2018towards,tao2020computing}: their privacy analysis requires an understanding of how the stability properties of two tables interact. Tumult Core handles this by defining a new metric that defines distances on sets of tables.
  \item \textbf{Complex neighboring notions.} Some datasets can contain an arbitrary number of records per user, which are keyed using a user ID. Tumult Core handles this by defining a metric for tables with privacy IDs, extending transformations to work with this metric, and adding truncation operators to convert from this metric to a metric quantifying the number of added/removed rows. Other complex neighboring relations are also implemented. %Other datasets can contain a fixed number of rows per user, but with some structural constraints on these rows; for example, that they all have distinct values in a specific column. This can be seen as a result of truncation on a table with privacy IDs, so the previous metrics can be reused.
  \item \textbf{Generalized parallel composition.} When running DP measurements on disjoint sets of records, we can reuse privacy budget, since each record can appear in at most one of these sets~\cite{mcsherry2009privacy}. Generalized parallel composition generalizes this result to non-disjoint subsets where a user's contribution across subsets is bounded. Tumult Core captures this contribution constraint in a metric between lists of datasets. %This allows for complex measurements where we (1) perform some transformation creating a list of datasets, with a constraint on a user's contribution across datasets; (2) remember this contribution constraint, in the form of a metric, while further transforming the data; (3) perform a measurement on each dataset, taking advantage of the contribution constraint to use privacy budget optimally.%
  %\footnote{This privacy accounting technique was used in a Tumult Analytics algorithm with similar structure for the decennial Census race and ethnicity product \todo{fix this, what exactly is it called?}. This allows for a significantly lower bound on the privacy loss than if we used less sophisticated accounting.}.
  \item \textbf{Adaptivity} Tumult Core supports fully adaptive mechanisms: when performing multiple measurements on the data, each measurement can reuse the output of previous measurements, including to determine how much privacy budget it consumes.
\end{itemize}

The variety of privacy definitions and accounting techniques can easily be combined. % due to the modular structure of Tumult Core.
Adding new notions of indistinguishability is as simple as defining a new measurement that supports this notion, or adding support to an existing measurement.
Then, the new measurement can be combined with existing transformations to produce complex mechanisms using the new privacy guarantee.
Similarly, new transformations that support e.g. privacy IDs can be combined with transformations that support more typical privacy accounting.
This way, new privacy accounting techniques can be used only when necessary, and existing components can be reused where it is not.
This modularity makes it easy to extend Tumult Core with new desired features. %, or improve as they are needed, or in response to new research in the academic community.
%In fact, the advanced features described above were added to Tumult Core in response to requirements from real use cases: the framework was not initially designed to support them.

\section{Tumult Analytics}
\label{sec:analytics}

While Tumult Core is powerful, its extremely modular design can make it challenging to use directly, especially for users who are not experts in differential privacy.
This is why we built Tumult Analytics as an interface layer on top of Tumult Core.
Tumult Analytics is designed to be usable for non-experts, while still being capable of addressing complex practical use cases.
We first discuss its design and why we believe it is approachable, using Figure~\ref{fig:analytics-example} as an example.
Second, we present some current real-world use cases of Tumult Analytics to give evidence of its power.

Computations in Tumult Analytics happen in the context of a \emph{Session}, which associates a fixed privacy guarantee with a dataset (possibly containing multiple tables), and mediates the full analysis run on this data.
An analyst defines the privacy budget when they create a Session, and they can query the Session for how much budget is remaining at any time.
They also define the privacy definition at Session creation -- both the indistinguishability definition (e.g. pure DP or zCDP) and the unit of privacy (e.g. how many records a single user contributes to in the original dataset).
Session creation is the only time the analyst must provide private data.
All further interactions between the analyst and the private data happen through the Session, preventing inadvertent privacy violations that may result from the analyst using the private data inappropriately.
The syntax for initializing a Session is demonstrated in the first block of Figure~\ref{fig:analytics-example}.

Next, the analyst can define \emph{queries}.
The query language attempts to limit the amount of new concepts specific to DP that the user has to learn and specify\footnote{Some, like noise type or magnitude, are either optional or entirely hidden. Others, like clamping bounds or explicit group-by keys, are required today, but work is currently underway to remove the need for users to understand and specify these.}.
This allows the analyst who is unfamiliar with DP to specify the statistics they want to see, and defers the process of constructing a private mechanism that approximates the answers to these queries to the Tumult Analytics engine.
This query construction process uses a fluent interface and generally attempts to resemble the pandas~\cite{reback2020pandas} and PySpark~\cite{pyspark} query interfaces, which users might already be familiar with.
This syntax is shown on the second block of Figure~\ref{fig:analytics-example}.

Once a query has been defined, the analyst evaluates it to produce a noisy answer.
This consumes privacy budget: the analyst must specify how much budget to use the query as in the third block of Figure~\ref{fig:analytics-example}.
When the analyst evaluates a query, Tumult Analytics compiles it into a Tumult Core measurement that answers the query using the given privacy budget, and returns the answer to the user.

Because Tumult Core supports fully adaptive composition, the entire process is interactive: the user can process the query answers and add complex control flow (conditional statements, loops, etc.) to choose which queries to evaluate next.
In the fourth block of Figure~\ref{fig:analytics-example}, the user asks the Session how much budget it has left.
The answer is greater than 0, so the user can run further queries.

To help users ramp up with DP, Tumult Analytics also comes with extensive documentation~\cite{analyticsdocs} and multiple tutorials~\cite{analyticstutorials} explaining how to perform common DP analysis tasks.
%We note that while we have received positive feedback on the interface and its onboarding materials, we have not yet performed formal user studies to experimentally validate this.
%We leave this to future work.

Ease-of-use is only valuable if users can actually solve their real-world data analysis problems.
We believe that Tumult Analytics is sufficiently powerful for such complex tasks.
The features discussed in Section~\ref{sec:core} are either available or currently being added, including private joins, multiple privacy definitions (pure differential privacy and zCDP), and support for privacy identifiers.
Tumult Analytics has already been deployed in practice, either in production or as a prototype in the following cases.
\begin{itemize}
  \item The U.S. Census Bureau is using Tumult Analytics for the Detailed DHC-A, Detailed DHC-B, and S-DHC data releases, as part of the 2020 Decennial Census. These use cases rely on several complex features outlined in Section~\ref{sec:core}: zCDP accounting, adaptivity, and tight privacy accounting with generalized parallel composition.
  \item The Wikimedia Foundation uses it to publish country-level statistics about the number of visitors to each Wikipedia page on each day. This use case relies on the complex privacy notion feature mentioned in Section~\ref{sec:core}.
  \item The U.S. Internal Revenue Service relies on Tumult Analytics to release college graduate income summaries and power the College Scorecard website~\cite{collegescorecard}.
\end{itemize}

%\section{Performance}
%\label{sec:performance}

%Tumult Analytics was built with performance in mind.
%To scale to large datasets, all operators in Tumult Core are implemented on top of Spark.
%Although answering queries using provable privacy adds some overhead that is not present when using PySpark directly, the result is still a system capable of scaling to the very large datasets.
%For example, the Census use case discussed in the previous records has \todo{number here} rows in the input, and outputs \todo{number} statistics.
%The Wikimedia data release has \todo{number} rows in the input, and outputs \todo{number} statistics.
%Both pipelines complete in a few hours on the Spark clusters of these organizations.
%
%\dd{I'm not a big fan of this section, it feels short and hand-wavy, especially the Spark cluster size part.}
%
%\dd{Should we add a paragraph to this section about security, referencing the work we've done securiting primitives?}

% \item In synthetic benchmarks of a simple groupby-count query, we show that the differentially private version of the query evaluated on the \tp scales similarly to the analogous pyspark query \todo{add figure or benchmark results, or remove this bullet}.

\section{Comparison to existing systems}
\label{sec:comparison}

Tumult Analytics is not the first open-source framework for differential privacy. In this section, we list other libraries, and explain how Tumult Analytics compares to them.

GoogleDP~\cite{googledp} is a suite of tools including ``building block'' libraries and higher-level frameworks (Privacy on Beam~\cite{privacyonbeam} written in Go, and an extension to ZetaSQL~\cite{zetasql,wilson2020differentially}).
Besides the obvious language difference (Tumult Analytics is a Python library), a main difference between Tumult Core and the GoogleDP building block libraries is that the latter do not provide an end-to-end guarantee: higher-level frameworks have to use them correctly and implement privacy-critical operations like privacy accounting or contribution bounding directly.
Another fundamental difference is extensibility.
GoogleDP tools are designed to support a specific class of queries~\cite{wilson2020differentially}, and use approximate DP for privacy accounting.
Extending them to support some of Tumult Analytics' features (like zCDP accounting, parallel composition, or private joins without privacy IDs) would require deep changes throughout the framework and the building block libraries.
By comparison, such changes were added to Tumult Core and Tumult Analytics in a matter of weeks when the need arose.

PipelineDP~\cite{pipelinedp} is a Python framework built atop the GoogleDP building block libraries.
It follows roughly the same design as the two GoogleDP frameworks, so the comparison from the previous paragraph applies here as well.
One major difference is that PipelineDP's design is backend-agnostic: it can run on multiple data processing frameworks, like Beam, Spark, or locally.
This is an advantage over Tumult Analytics, which fully relies on Spark.

OpenDP%
\footnote{The software library~\cite{opendp}, not to be confused with the larger project with the same name~\cite{opendpproject}.
Another software library under the same project is SmartNoise Core~\cite{smartnoisecore}, which is an older version of OpenDP and is now deprecated.}~\cite{opendp}.
is inspired by a programming framework proposed by \cite{gaboardi2020programming}.
This framework was also the inspiration for Tumult Core, 
so there are similarities between both projects.
Besides feature-richness (the features mentioned in Section~\ref{sec:core} do not exist yet in OpenDP), a major difference between the two projects is scalability: all components in OpenDP are written in Rust, so transformations like group-by operations and aggregations assume that all the data fits in memory on a single machine.
This makes it unsuitable for large-scale data processing use cases.

SmartNoise SQL~\cite{smartnoisesql} is a high-level framework to run differentially private SQL queries. 
It uses some primitives from OpenDP, like noise addition, but (maybe due to the scalability limitations mentioned above) uses the SQL backend directly for most data processing operations.
As a result, most of the privacy-critical logic is implemented directly in SmartNoise SQL instead of using the underlying framework as was originally envisioned~\cite{gaboardi2020programming}. 
This means that even though OpenDP can generate a proof of privacy for a complex program from its simple components, SmartNoise SQL cannot.
This architectural choice also means that SmartNoise SQL cannot directly benefit from OpenDP's extensibility.

Diffprivlib~\cite{diffprivlib} is a library of DP mechanisms written in Python.
It relies on NumPy~\cite{numpy,harris2020array} for all the underlying computations, which has the same scalability limits as OpenDP, and also creates security vulnerabilities~\cite{haney2022precision}.
diffprivlib is also not built on an extensible framework like OpenDP or Tumult Core; it relies exclusively on approximate DP, and can only protect individual records in a single table.

Other systems have been proposed in the literature, and proofs of concept for these systems have been published on open-source platforms: this is the case for Chorus~\cite{chorus,johnson2020chorus} or PINQ~\cite{pinq,mcsherry2009privacy}.
Since these libraries are prototypes that are not actively maintained, we do not extensively compare Tumult Analytics to them.

%\vspace{\baselineskip}
\section{Acknowledgments}

We are thankful to Jahlela Hasle, Michael Fine, Kiron Lebeck, Pira Limpiti, David Pujol, Simran Rajpal, and Daniel Simmons-Marengo for their contributions to Tumult Analytics and helpful feedback on this paper.

\bibliographystyle{IEEEtran}
\bibliography{ref}

\end{document}